\documentclass[notitlepage,a4,10.5pt]{article}%,twocolumn]{revtex4}

\usepackage{graphicx}

\usepackage{amsmath}%
\usepackage{amsfonts}%
\usepackage{amssymb}%
\usepackage{mathrsfs}
\usepackage{amsthm}
\usepackage{float}
\usepackage[numbers,sort&compress]{natbib}

\usepackage{authblk}

\usepackage[left=2.5cm,right=2.5cm,top=3cm,bottom=3cm]{geometry} 

\usepackage{xcolor}

\usepackage{newtxtext}
\usepackage{newtxmath}
\usepackage{natbib}
\usepackage{hyperref}

\usepackage{array}
\usepackage{colortbl} % For colored rows and table rules
\usepackage{graphicx} % For better compatibility with figures (optional)

\newcommand{\cp}{\times}

\newcommand{\Norm}[1]{\left\lvert\left\lvert #1 \right\rvert\right\rvert}

\newcommand{\bol}{\boldsymbol}

\newcommand{\abs}[1]{\left\lvert{#1}\right\rvert}

\newcommand{\lr}[1]{\left({#1}\right)}
\newcommand{\lrs}[1]{\left[{#1}\right]}
\newcommand{\lrc}[1]{\left\{{#1}\right\}}

\newcommand{\p}{\partial}

\newcommand{\ti}[1]{\textit{#1}}

\newcommand{\eq}[1]{\begin{equation}\begin{split}{#1}\end{split}\end{equation}}
\newcommand{\sys}[2]{\begin{subequations}\begin{align}{#1}\end{align}\label{#2}\end{subequations}}

\begin{document}

\title{Axisymmetric Dynamos Sustained by a Modified Ohm's Law in a Toroidal Volume}
\author[1]{Naoki Sato} \author[1]{Kumiko Hori}
\affil[1]{National Institute for Fusion Science, \protect\\ 
322-6 Oroshi-cho, Toki-city, Gifu 509-5292, Japan \protect\\ Email: sato.naoki@nifs.ac.jp}
\date{\today}
\setcounter{Maxaffil}{0}
\renewcommand\Affilfont{\itshape\small}

%\twocolumn[
 % \begin{@twocolumnfalse}
    \maketitle
    \begin{abstract}
This work tackles a significant challenge in dynamo theory: the possibility of long-term amplification and maintenance of an axisymmetric magnetic field. We introduce a novel model that allows for non-trivial axially-symmetric steady-state solutions for the magnetic field, particularly when the dynamo operates primarily within a ``nearly-spherical'' toroidal volume inside a fluid shell surrounding a solid core. 
In this model, Ohm's law is generalized to include the dissipative force, arising from electron collisions, that tends to align the velocity of the shell with the rotational speed of the inner core and outer mantle.
Our findings reveal that, in this context, Cowling's theorem and the neutral point argument are modified, leading to magnetic energy growth for a suitable choice of toroidal flow. The global equilibrium magnetic field that emerges from our model exhibits a dipolar character. 
The central insight of the model developed here is that 
if an additional force is incorporated into Ohm's law, symmetric dynamos become possible.	
\end{abstract}
%\vspace{5mm}
%\end{@twocolumnfalse}
%  ]

\section{Introduction} \label{sec:intro}

Dynamo theory posits that planetary magnetic fields are generated by converting kinetic energy into magnetic energy via Faraday's law of induction in a magnetohydrodynamic (MHD) fluid \cite{Moffatt78,Desjardins07,Tobias21}.
The mechanism behind the long-term amplification and maintenance of these fields, particularly axisymmetric magnetic fields, remains a longstanding theoretical problem. 
This inquiry dates back to \cite{Larmor19}, who hypothesized that a flow of ionized gas (as in the Sun) occurring in a magnetic field aligned with an axis would induce a toroidal electric current circulating about the axis, thereby sequentially enhancing the magnetic field. However, Cowling's neutral point argument \cite{Cowling33} established the well-known antidynamo theorem, stating that an axisymmetric magnetic field cannot be sustained by dynamo action.  
This finding laid the groundwork for mean-field theory \cite{Krause80} 
% ,Roberts71
and nearly-axisymmetric dynamos \cite{Brag76}, while kinematic approaches explored magnetic field enhancement through predefined velocity fields \cite{Backus58,Ponomarenko73}. 
%For comprehensive overviews, we refer to these references. 
Meanwhile, theoretical efforts aimed to generalize Cowling's theorem by removing the restrictions noted in his original proof and possibly finding ways to circumvent it (see detailed discussions in \cite{Ivers84}, \cite{Nunez96}, and \cite{Kaiser14}). Allowing anisotropy in conductivity emerged as a potential solution to support symmetric field configurations \cite{Lortz89,Plunian20}. 
Interestingly, modeling variable conductivity in gaseous planets,  
%like Saturn, 
state-of-the-art numerical MHD dynamo simulations exhibited a quasi-equilibrium field that was almost-axisymmetric and dipole-dominated \cite{Yadav22} . 

In this study, we revisit the fundamental question of whether axisymmetric dynamos can be theoretically realized within the framework of kinematic dynamo theory. 
Here, an axisymmetric dynamo refers to the spontaneous generation of a magnetic field that exhibits axial symmetry. Our hypothesis is that the resistive force arising from electron-electron collisions—commonly neglected in standard magnetohydrodynamics (MHD)—may play a  role in planetary dynamos. 
We highlight that this MHD regime serves as a simplified model for certain planetary interiors, as such systems—whether consisting of fully ionized plasma or liquid metal—can be effectively modeled as a two-fluid system of ions and electrons interacting through electromagnetic forces. The resistive force due to electron-electron collisions becomes particularly significant when the turbulence scale of the fluid flow in planetary interiors is sufficiently small, thereby requiring a modification to Ohm's law. Indeed, the constitutive relation connecting the electric field to the magnetic field and velocity field depends on the forces acting on the electron fluid \cite{Freid}. 
To explore this idea, we propose two simplified models for the resistive force: a restoring friction force model and a viscous dissipation model. Importantly, the emphasis of this work lies in the underlying physical principle: \textit{if an additional force is incorporated into Ohm's law, symmetric dynamos become possible.}

In fusion plasma physics, stable equilibrium magnetic fields are crucial for developing magnetic confinement fusion reactors \cite{Wesson04,Helander14}. A key property of fusion reactors is that they must be toroidal rather than spherical, due to the impossibility of accommodating a non-vanishing magnetic field on a spherical surface (hairy-ball theorem; \cite{Eisenberg79}). 
%Additionally, Ohm's law, which relates the electric field to the magnetic field and velocity field, varies depending on the forces acting on the electron fluid \cite{Freid}. 
%This work leverages these principles by 
%modeling the planetary dynamo-active region located between the inner core and the outer mantle as a ``nearly-spherical'' thick toroidal volume nested within the 
%spherical shell (the volume one would obtain by attempting to fill a hollow sphere with a torus, see section 2 and fig. 1 therein) 
%and by applying a generalized Ohm's law that incorporates the dissipative force arising from electron collisions. 
This work builds on these principles by modeling the planetary dynamo-active region, situated between the inner core and the outer mantle, as a `nearly-spherical' thick toroidal volume enclosed within the spherical liquid metal  shell. This volume resembles the shape formed when attempting to fit a torus into a hollow sphere (see Section 2 and Figure 1). Applying the modified Ohm's law that accounts for the resistive force caused by electron collisions, we show that the resulting dynamo model successfully generates an axially symmetric dipolar magnetic field driven by an ion flow aligned with the steady electric current density. We also uncover that, at equilibrium, the ideal convective (Lorentz) forces and the dissipative (Ohmic resistance and friction/viscosity) forces balance independently. This regime aligns with the kinetic theory of plasmas \cite{Sato24}, from which magnetohydrodynamics (MHD) originates.
%this dissipative force can be comparable to the conventional resistive force, associated with electron-ion collisions in the standard formulation of Ohm's law. Therefore, it should not be neglected.

The present model may thus serve as an illustrative example for understanding the dynamo mechanism responsible for generating the dominant dipole component of planetary magnetic fields, including the highly symmetric configurations observed in 
Saturn and Mercury \cite{Cao23,Anderson12}.   
The nearly axisymmetric fields of the gas giant and the rocky planet have  been attributed to the influence of a stably stratified layer above the dynamo region \cite{Stevenson80,Christensen06}. Our model suggests that this can naturally result from dynamo action confined within a toroidal volume inside the spherical dynamo region, 
influenced by the balance of resistive and friction/viscosity forces.

\section{Modeling an internal dynamo  region}\label{sec:Equilibrium}

%In this section we provide a basic modeling of the dynamo  of the Earth or any other planet 
%(or any Earth-like planet) 
%by characterizing the topology of the interior regions below its surface. 

%In this section, we present a foundational model for the dynamo mechanism of Earth or any other planet by characterizing the topology of the interior regions beneath its surface. 
%This step is essential for achieving dynamo action, as the toroidal topology of the current density flow is critical for the existence of non-trivial solutions to the induction equation. Specifically, identifying a toroidal domain for the current flow allows us to establish the appropriate boundary conditions for the magnetic flux, which are essential for the existence of solutions that would not be possible in a purely spherical geometry.
%This step is crucial to achieving dynamo action, because it is the toroidal topology of the current density flow that allows non-trivial solutions of the induction equation  to exist. More precisely, the identification of a toroidal domain for the flow of current enables the identification of the appropriate boundary conditions for the magnetic flux, without which solutions would not exist. 

In this section, we introduce a simplified model of the interior regions beneath the surface of a planet with volume \( S \subset \mathbb{R}^3 \). Specifically, we model the dynamo-active region \( T \) — the region where the electric current density responsible for dynamo action is nonvanishing — as a toroidal volume \( T \subset S \).

The choice of a toroidal volume is motivated by the nature of axially symmetric dipole magnetic fields generated by spontaneous dynamo action. In such cases, the corresponding electric current density is toroidal and can be expressed as \( \bol{J} = \alpha(r,z)\nabla\varphi \), where \( (r,\varphi,z) \) are cylindrical coordinates, and \( \alpha(r,z) \) is a function of \( r \) and \( z \). Due to the singularity of \( |\nabla\varphi| = 1/r \) at \( r=0 \), 
and the fact that the direction of  $\nabla\varphi$ is not uniquely defined at $r=0$, 
the current density \( \bol{J} \) must vanish along the planetary axis (\( r=0 \)). Consequently, a dynamo-active region sustaining such an an axially symmetric magnetic field cannot have a spherical topology. Instead, it can be appropriately modeled as a toroidal volume \( T \subset S \) embedded within the planetary volume \( S \subset \mathbb{R}^3 \).

This observation is closely related to the Poincaré-Hopf and hairy ball theorems \cite{Eisenberg79}, which state that a non-vanishing continuous vector field cannot exist tangent to a spherical surface. For a sphere to accommodate a toroidal current, there must be regions where the current vanishes. Practically, this implies the absence of significant macroscopic currents near the planet's poles, supporting the choice of a topologically toroidal dynamo-active region.

We further assume that the magnetic axis aligns approximately with the planet's rotational axis and that \( T \) exhibits rotational symmetry around the vertical axis connecting the poles. The planetary structure is modeled with a solid core \( C \), surrounded by a fluid shell \( L \), which lies beneath the mantle \( M \). Here, \( C \) is a spherical volume, while \( L \) and \( M \) are hollow spherical shells. Both the solid core and fluid shell may be electrically conducting; however, the solid nature of the core suggests that convective currents sustaining the planet’s magnetic field are confined to the fluid shell. Hence, \( T \subset L \).

Additionally, we assume that the toroidal volume \( T \) nearly fills the fluid shell \( L \), resulting in a `nearly-spherical' yet topologically toroidal region (see Fig. 1). For the purposes of this study, we focus on \( T \) and the exterior region \( O = \mathbb{R}^3 \setminus \bar{T} \), where \( \bar{T} \) denotes the closure of \( T \).

A schematic representation of this model is shown in Fig. 1. In this figure, \( \bol{J}_{\infty}(\bol{x}) = \mu_0^{-1} \nabla \times \bol{B}_{\infty}(\bol{x}) \) and \( \bol{B}_{\infty}(\bol{x}) \) denote the long-term current density and magnetic field, respectively, representing the final equilibrium fields resulting from dynamo action. Here, \( \mu_0 \) is the vacuum permeability.

%and that the toroidal volume $T$ almost fills  
%an Earth-like planet, 
%which we model in terms of a spherical volume $S\subset\mathbb{R}^3$. 
%We note, however, that the same construction is applicable to any toroidal domain \( T \), including those with asymmetric configurations. 
%Another way to deduce that the domain \( T \) hosting the current density flow should be toroidal, particularly when the flow is predominantly toroidal, involves the Poincaré-Hopf and hairy ball theorems \cite{Eisenberg79}. These theorems state that there can be no non-vanishing continuous vector field tangent to a spherical surface. Consequently, accommodating a toroidal current on a sphere necessitates points where the flow vanishes. Practically, this implies that we should not expect macroscopic currents in the regions surrounding the planet's
%planetary 
%poles.

%The structure of the deeper interior region of the planet can be assessed from indirect evidence based on the values of pressure, temperature, and density: for the Earth, 

%%%%%%%%%%%%%%%%%%%%%%%%%%
\begin{figure}
  \centerline{\includegraphics[scale=0.2]{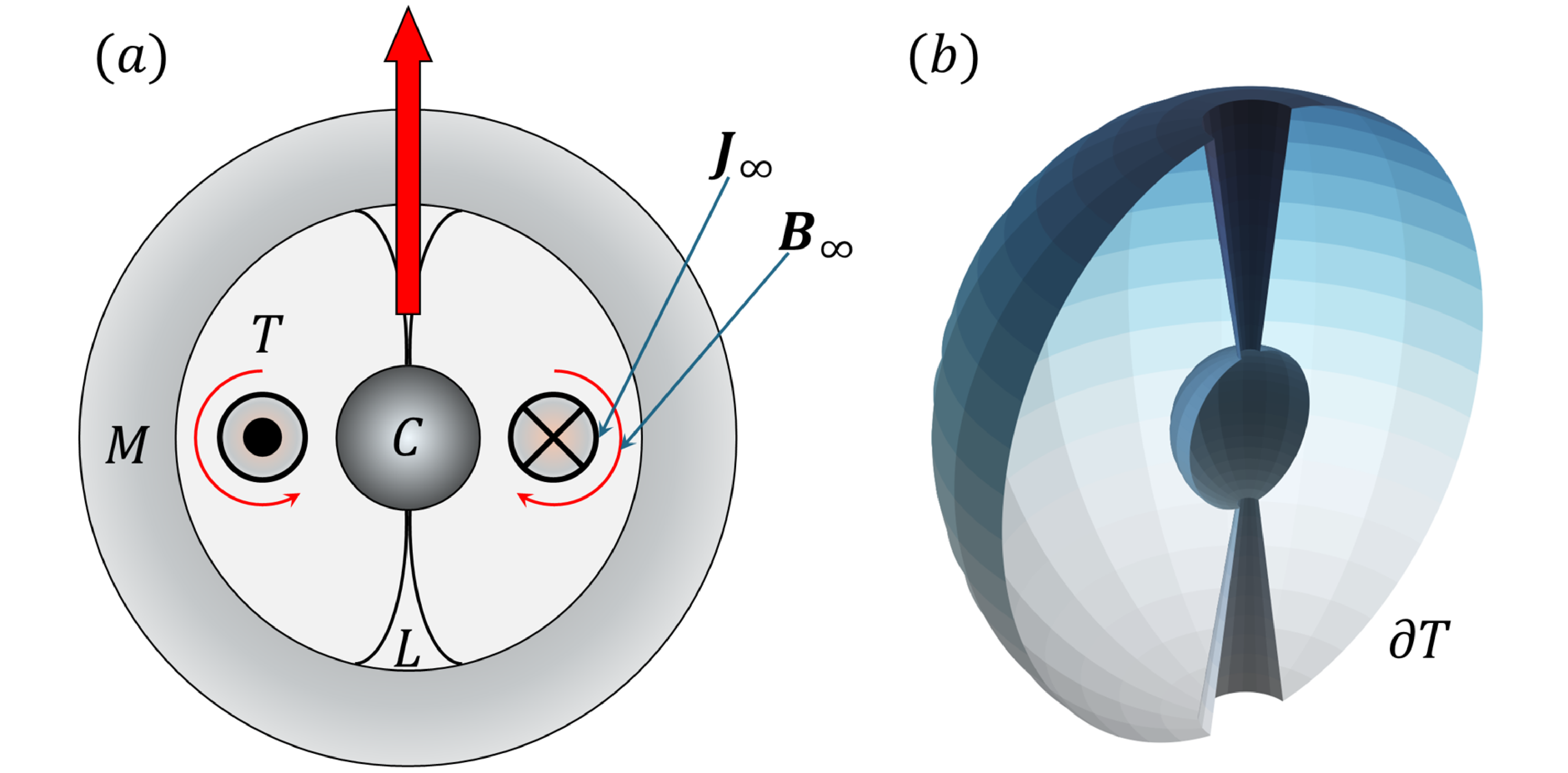}}% Images in 100% size
  \caption{ (a) Schematic representation of a dynamo region in a planet.
  %the Earth's interior regions. 
  Note that the current density $\bol{J}_{\infty}$ responsible for the dipole magnetic field $\bol{B}_{\infty}$ flows within a toroidal volume $T$ entirely contained within the fluid 
  %liquid 
  shell $L$. (b) The half-cut  toroidal surface $\p T$ bounding the the toroidal dynamo region $T$. Note that the current density \( \boldsymbol{J}_{\infty} \) vanishes in the dark region surrounding the vertical axis, which corresponds to the toroidal hole.}
\label{fig:ka}
\end{figure}
%%%%%%%%%%%%%%%%%%%%%%%%%%

\section{Ohm's Law in a Rotating Viscous Fluid}
In the following, we adopt a simplified model of the electrically conducting fluid responsible for dynamo action, described as an ion-electron two-fluid system. This simplified approach is not intended for making quantitative predictions about the magnetic properties of specific planets or stars. Instead, it provides a conceptual framework that highlights the core physical principle: \textit{if an additional force is incorporated into Ohm's law, symmetric dynamos become possible.}

Ohm's law relates the electric field $\bol{E}\lr{\bol{x},t}$ to the velocity field $\bol{u}\lr{\bol{x},t}$, the magnetic field $\bol{B}\lr{\bol{x},t}$, and the current density $\bol{J}\lr{\bol{x},t}$ through the resistivity $\eta>0$, modeled as a constant, according to
\eq{
\bol{E}=\bol{B}\cp\bol{u}+\eta\bol{J}.\label{Ohm}
}
Equation \eqref{Ohm} originates from the electron fluid momentum equation
\eq{
m_en_e\frac{d\bol{u}_e}{dt}=-en_e\lr{\bol{E}+\bol{u}_e\cp\bol{B}}-\nabla P_e+\bol{R}_e,\label{due}
}
where $m_e$, $-e$, $n_e\lr{\bol{x},t}$,  $\bol{u}_e\lr{\bol{x},t}$, $P_e\lr{\bol{x},t}$, and $\bol{R}_e\lr{\bol{x},t}$ denote the electron mass, charge, number density, fluid velocity, pressure, and resistive force resulting from ion-electron collisions   respectively \cite{Freid}. 
Recalling that the electron fluid velocity $\bol{u}_e$ is related to the current density $\bol{J}$, the ion fluid velocity $\bol{u}_i\lr{\bol{x},t}$, the ion number density $n_i\lr{\bol{x},t}$, and the ion charge $Ze$  according to  
$%\eq{
\bol{J}=e\lr{Zn_i\bol{u}_i-n_e\bol{u}_e}$,
%}
equation \eqref{due} can be solved for $\bol{E}$. %, 
%\eq{
%\bol{E}=Z\frac{n_i}{n_e}\bol{B}\cp\bol{u}_i+\frac{1}{en_e}\bol{J}\cp\bol{B}-\frac{1}{en_e}\nabla P_e+\frac{\bol{R}_e}{en_e}-\frac{m_e}{e}\frac{d\bol{u}_e}{dt}.
%}
Due to the smallness of the electron to ion mass ratio $\delta=m_e/m_i$, 
the velocity of the center of mass of the ion-electron $2$-fluid system 
effectively corresponds to the ion fluid velocity, i.e. $\bol{u}=\frac{\bol{u}_i+\delta\bol{u}_e}{1+\delta}\approx\bol{u}_i$. Furthermore, assuming quasi-neutrality, we have $n\lr{\bol{x},t} =n_i\approx n_e/Z$, leading to
\eq{
\bol{E}=\bol{B}\cp\bol{u}+\frac{1}{Ze n}\bol{J}\cp\bol{B}-\frac{1}{Zen}\nabla P_e+\frac{\bol{R}_e}{Zen}-\frac{m_e}{e}\frac{d\bol{u}_e}{dt}.\label{Ohm2}
}
The second term on the right-hand side is the Hall
effect. In the regime of magnetohydrodynamics (MHD), both the Hall and electron pressure terms \eqref{Ohm2} scale with  the ion-gyroradius, and are therefore neglected on the larger spatial scale of the system.  
The last term on the right-hand side of \eqref{Ohm2} is similarly neglected under the standard MHD assumption \cite{Freid} that the small electron mass %\naoki{and the slow temporal variation of $\bol{u}_e$}  
makes it smaller than all the other terms, 
including the resistive term $\bol{R}_e/Zen$. If we further assume a linearity relationship between $\bol{R}_e$ and the current density $\bol{J}$, i.e. \eq{\frac{\bol{R}_{e}}{Zen}=\eta\bol{J},\label{Re}} we arrive at Ohm's law \eqref{Ohm}. 
For completeness, it is important to note that the derivation of Ohm's law presented above aligns well with the classical Drude model \cite{Ash}, which describes electrons as following random trajectories due to collisions with stationary ions. The emergence of a steady-state characterized by a linear response to the applied electric field is intrinsically tied to the small mass of the electron. This small mass enables an exceptionally fast response time relative to the MHD time scale, thereby ensuring that the system quickly reaches equilibrium where the current density is proportional to the electric field.

The standard form of Ohm's law \eqref{Ohm} is valid only within the parameter regime of magnetohydrodynamics (MHD). However, it is well-established that Ohm's law must be modified when the assumptions underlying this regime no longer hold. For a detailed discussion on the applicability of MHD, we refer to \cite{Freid} and \cite{Fitz}. Generalized forms of Ohm's law, such as those in Hall MHD \cite{Ach} and extended MHD \cite{Abd}, have been employed in applications to dynamo theory \cite{Minnini,Lingam}. The central idea of this study is that the resistive force arising from electron-electron collisions, which is typically neglected in standard MHD, may play a significant role in planetary dynamos. This necessitates a corresponding modification of Ohm's law \eqref{Ohm}. In the remainder of this section, we develop a model for this modified form of Ohm's law.

%In the absence of a magnetic field, Ohm's law \eqref{Ohm} implies that
%the electric field is proportional to the current density,  
%$\bol{E}=\eta\bol{J}$. 
Now suppose that we move to a rotating reference frame with speed $\bol{v}_R=\frac{1}{2}\Omega r^2\nabla\varphi$, where $\Omega>0$ is a constant, and $\lr{r,\varphi,z}$ is a cylindrical coordinate system. In the rotating frame, the ion and electron fluid velocities become $\bol{u}'=\bol{u}-\bol{v}_R$ and $\bol{u}_e'=\bol{u}_e-\bol{v}_R$. It follows that
\eq{
\bol{J}=Ze n\lr{\bol{u}-\bol{u}_e}=Zen\lr{\bol{u}'-\bol{u}_e'}=\bol{J}',\label{EOhm}
}
where {the prime} 
%the $'$ symbol 
means that the quantity is evaluated in the rotating frame. %Equation \eqref{EOhm} is telling us that the electric field in the rest frame and the electric field in the rotating frame remain unchanged. 
%On the other hand, if we express the electron momentum equation in the rotating frame, neglecting Hall effect, pressure force, acceleration $d\bol{u}_e'/dt$, and centrifugal force (which scales with the square of $\Omega$), we obtain
%\eq{
%\bol{E}'=\bol{B}\cp\bol{u}'+\frac{\bol{R}_e'}{Zen}+\frac{m_e}{e}\Omega\,\bol{u}'\cp\nabla z=\bol{B}\cp\bol{u}+\frac{\bol{R}'_e}{Zen}+\frac{m_e}{e}\Omega\,\bol{u}\cp\nabla z,%-\frac{m_e}{e}\frac{d\bol{u}_e'}{dt}, 
%}
%where we used the fact that $\bol{B}=\bol{B}'$ and neglected terms quadratic in $\Omega$. 
%Since in the absence of a magnetic field the electric field in the rest frame $\bol{E}$ an the electric field in the rotating frame $\bol{E}'$ must coincide, at leading order in $\Omega$ we find 
%\eq{
%\bol{R}_e'=\bol{R}_e-\frac{m_e}{e}\Omega\,\bol{u}\cp\nabla z=\eta\bol{J}-\frac{m_e}{e}\Omega\,\bol{u}\cp\nabla z.\label{Re2}
%}
Equations \eqref{Re} and \eqref{EOhm} show that the resistive force $\bol{R}_e$ %in the electron momentum equation 
cannot  take into account dissipative forces  %associated with the macroscopic rotation $\bol{v}_R$ of the system
that cannot be expressed through the relative velocity $\bol{u}-\bol{u}_e$. Indeed, the force $\bol{R}_e$ does not change in the presence of rotation $\bol{v}_R$ due to the invariance of $\bol{J}$ described by \eqref{EOhm}.  
However, the fluid shell surrounding the {solid} core cannot slide: it is subject to friction forces that tend to uniform the speed of the fluid with the rotation speed $\bol{v}_R$ of {the inner 
%solid 
core and the overlying  
%outer 
mantle. 
Now imagine sliding a condensed viscous fluid on a rough surface. 
It is then unlikely that these friction forces can be neglected in the modeling of the fluid shell.}
%Due to the extreme pressure,  temperature, and density conditions in planetary interiors, 
{The simplest model for 
this type of friction %can be modeled in terms of 
is given by a restoring force $\bol{F}_{\gamma}$ acting on the electron fluid}
\eq{
\bol{F}_{\gamma}=-Ze\gamma n\lr{\bol{u}_e-\bol{v}_R}=-Ze\gamma n\lr{\bol{u}-\bol{v}_R}+\gamma\bol{J},\label{Fgamma}
}
where $\gamma>0$ is a physical constant.  %(in eq. \eqref{Fgamma}, the ion charge $Ze$ is put in front of $\gamma$ for  later convenience). 
The result is a modified Ohm's law, 
\eq{
\bol{E}=%\bol{B}\cp\bol{u}+\lr{\eta-\frac{\gamma}{Zen}}\bol{J}-\gamma\lr{\bol{u}-\bol{v}_R}=
\bol{B}\cp\bol{u}+\eta_{\gamma}\bol{J}-\gamma\lr{\bol{u}-\bol{v}_R},~~~~\eta_\gamma=\eta+\frac{\gamma}{Zen}.\label{GOhm}
}
Note that we introduced an effective resistivity $\eta_{\gamma}$.  
Remarkably, the restoring force contains a toroidal component $\lr{\bol{u}-\bol{v}_{R}}\cdot\nabla\varphi$ that can potentially balance a toroidal current density $\bol{J}\cdot\nabla\varphi\neq 0$. 
Indeed, the difference $\bol{u}-\bol{u}_e$ between ion fluid velocity and electron fluid velocity can have the same orientation as the difference $\bol{u}_e-\bol{v}_{R}$ between electron fluid velocity and rotation velocity, implying that 
the resistive force $\eta\bol{J}$ can be balanced by the restoring force $-\gamma\lr{\bol{u}_e-\bol{v}_{R}}$. 
Physically, the fast moving ions attract the electron fluid in one direction, while rotation of the planet  effectively drags them in the other. 

{It is worth noting that the resistive force, which is proportional to $\bol{u} - \bol{u}_e$, and the frictional force, which is proportional to $\bol{u} - \bol{v}_{R}$, share the same mathematical structure. In the case of the resistive force, this structure represents the linear response to electron-ion collisions. Conversely, the frictional force models the linear response to electron-electron collisions, which interact with the rotating inner core and outer mantle as they collide with the boundary of the toroidal domain, propagating this effect through subsequent collisions.
}
We also remark   that the friction force $\boldsymbol{F}_{\gamma}$ is consistent with Galilean invariance, as the velocity differences $\boldsymbol{u} - \boldsymbol{v}_R$ and $\boldsymbol{u} - \boldsymbol{u}_e$ remain invariant across inertial frames under Galilean relativity.

%\naoki{UNDER CONSTRUCTION
We emphasize that \eqref{Fgamma} represents a simplified description of the actual friction force at play in the {fluid} shell, and is chosen to simplify the later analysis. 
This friction force is ultimately rooted in the momentum transfer associated with  electron collisions. 
{An accurate derivation of the friction force should take into account the individual condition of the planet,
 e.g.} the graduality of the transition from liquid to solid phase {in Earth's interior},
 and would require kinetic modeling of particle collisions in a rotating setting, leading to a corresponding electron fluid momentum equation subject to boundary conditions for the velocity field that take into account 
{the planet's rotation}. 
If viscous resistive MHD equations are chosen to model the electron momentum equation in the rotating frame, the resulting modification of Ohm's law in the non-rotating rest frame is
\eq{
\bol{E}=\bol{B}\cp\bol{u}+\eta\bol{J}+\frac{m_e\nu_e}{e}\Delta\bol{u}_e\approx \bol{B}\cp\bol{u}+\eta\bol{J}+\frac{m_e\nu_e}{e}\Delta\bol{u}, \label{GOhm2}
}
where $\nu_e$ is the electron fluid kinematic viscosity, $\Delta$ the Laplacian operator, we used the fact that $\Delta\bol{v}_R=\bol{0}$, and neglected the term proportional to $\Delta\lr{\bol{J}/Zen}$, which is expected to represent a higher order correction in the plasma regime under consideration. {Note that, although $\bol{v}_R$ does not appear explicitly in $\Delta\bol{u}$, the presence of the operator $\Delta$ implies that planetary rotation is felt throughout the toroidal domain via the no-slip boundary conditions on  $\bol{u}$.}

{It is} important to note that since the difference \(\bol{u} - \bol{u}_e\) can be small relative to \(\bol{u}\), the viscous term may be comparable to the resistive term. 
This implies that {
%the common practice of 
neglecting viscous effects in resistive MHD} may not always be a valid approximation, {as clear from the numerical examples given below}. In particular, note that, when approaching the boundary of the {fluid} shell, the current density progressively drops toward zero, leaving the friction/viscous forces the dominant toroidal component in Ohm's law.   
In fact, the condition $\eta\bol{J}\sim m_e\nu_e\Delta\bol{u}/e$ can be used to estimate 
%the order of the relative velocity $\bol{u}-\bol{v}_R$ 
the characteristic spatial scale $L$ of the relative velocity %$\bol{B}$ and 
$\bol{u}-\bol{v}_R$ in the {fluid} shell
needed to sustain a given current density $\bol{J}$. We have,
%\eq{
%\bol{u}-\bol{u}_e\sim\frac{\bol{J}}{en}\sim\frac{m_e\nu_e}{\eta ne^2L^2}\lr{\bol{u}-\bol{v}_R},
%}
%where $L$ the characteristic spatial scale of %$\bol{B}$ and 
%$\bol{u}-\bol{v}_R$ in the liquid shell. %Recalling that resistivity scales as $\eta\sim m_e/ne^2\tau_{ei}$ \cite{Freid}, where $\tau_{ei}$ denotes the electron-ion collision time, 
%It follows that
\eq{
\bol{u}-\bol{v}_R\sim \frac{eL^2\eta}{m_e\nu_e}\bol{J}\sim \frac{e\mu_0L^2}{m_e{\rm P_m}}\bol{J}  \quad 
\implies \quad 
L^2\sim \frac{m_e{\rm P_m}\abs{\bol{u}-\bol{v}_R}}{e\mu_0\abs{\bol{J}}},
%\frac{\bol{B}L}{\mu_0en\tau_{ei}\nu_e}\sim\frac{\bol{B}L}{\mu_0en\tau_{ei}\tau_{ee}V_e^2}%\sim 10^3 ms^{-1}
%\sim \frac{L}{\tau_{ee}^2} 10^{-17} s,
}
where 
%$L$ is the characteristic spatial scale of %$\bol{B}$ and 
%$\bol{u}-\bol{v}_R$ in the liquid shell and 
${\rm P_m}=\mu_0\nu_e/\eta$ is the (electron) magnetic Prandtl number.
For given $\bol{u}-\bol{v}_R$, we see that when 
%the magnetic Prandtl number 
${\rm P_m}$
is smaller
%, such as near the liquid shell-mantle interface, 
the current density and the turbulence scale $L$  tend to be very small. Conversely, in areas where 
%the magnetic Prandtl number 
${\rm P_m}$
is higher
%, such as near the solid core-liquid shell interface, 
substantial currents can flow and larger scales $L$ are allowed. 
As an example, consider {Earth's liquid iron core for} a relative velocity of \(\bol{u} - \bol{v}_R \approx 10^{-4} \, \text{m/s}\). 
% \cite{Gillet19}.
Using estimated values {for
%Earth's 
resistivity}, \(\eta \approx 40 \, \mu\Omega\,\text{cm}\) \cite{Ohta16}, and 
%liquid iron
kinematic viscosity, \(\nu \approx 1.5\times10^{-6} \, \text{m}^2/\text{s}\) \cite{Wijs98}, noting that $\nu_e\approx\sqrt{{m_i}/{m_e}}\nu$ \cite{Freid}, one finds a turbulence scale of \(L \approx 10^{-3}\text{m}\) for {the geo}magnetic field \(\bol{B} \approx 10^{-6} \, \text{T}\) such that \(\nabla \times \bol{B} \approx 10^{-13} \, \text{T/m}\).
%
%is comparable to \(L\), indicating that an accurate dynamo model for these parameters  should include the Hall effect and electron pressure in Ohm's law. Nevertheless, it can be shown that the theory developed in the present paper remains unchanged as these terms, which tend to align with the gradient of the magnetic flux \(\Psi\), can be balanced by a corresponding electric field. 
%
A similar estimate for {Saturn's metallic hydrogen, with $\nu\approx 0.4\times 10^{-6}{\rm m^2/s}$, $\eta\approx 0.2\times10^{-3} \mu\Omega\,\text{cm}$ \cite{Preising23}, 
%$\bol{B}\approx 10^{-6}\text{T}$,
and $\nabla\cp\bol{B}\approx 10^{-14}\text{T/m}$,
gives $L\approx 10\,\text{m}$ when $\bol{u}-\bol{v}_R\approx 10^{-2}\text{m/s}$.
%\cite{Bloxham24}.
%
%%%%%%%%%%%%%%%%%%%%%%%%%%
%For the Sun \cite{Jones10}, with $\bol{u}-\bol{v}_R\approx 10^2\text{m/s}$,  ${\rm P_m}\approx 10^{-3}\sqrt{m_i/m_e}$, $\bol{B}\approx 10^{-4}\text{T}$, and  $\nabla\cp\bol{B}\approx 10^{-13}\text{T/m}$, we find $L\approx 10\,\text{m}$ and $r_i\approx10^{-2}\text{m}$. 
%%%%%%%%%%%%%%%%%%%%%%%%%%
%
%Additionally, 
In those plasma regimes, by contrast, the ion gyroradii $r_{i}$ are found to be approximately \( 10^{-5} \, \text{m}\) and $10^{-4}\text{m}$ for the Earth and Saturn, respectively.
The MHD approximation with the modified Ohm's law is hence valid here (see Table 1). 
We also note that these estimates on $L$ likely represent lower bounds as $\nu_e\approx\sqrt{m_i/m_e}\nu$ only takes into account electron-electron collisions. }

\begin{table}%[h!]
\centering
\renewcommand{\arraystretch}{1.5}
\setlength{\arrayrulewidth}{0.2mm} % Thinner table borders
\setlength{\tabcolsep}{5pt} % Adjusts padding inside cells

\begin{tabular}{|>{\centering\arraybackslash}m{2.25cm}|>{\centering\arraybackslash}m{5cm}|>{\centering\arraybackslash}m{5cm}|}
\hline
\rowcolor{black} 
\textcolor{white}{\textbf{Parameter}} & \textcolor{white}{\textbf{Earth Liquid Iron Core}} & \textcolor{white}{\textbf{Saturn Metallic Hydrogen}} \\ \hline
\textbf{\(\boldsymbol{u} - \boldsymbol{v}_R\)} & \(10^{-4} \, \text{m/s}\) & \(10^{-2} \, \text{m/s}\) \\ \hline
\textbf{\({\eta}\)} & \(40 \, \mu\Omega \) (Ohta 2016) & \(0.2 \times 10^{-3} \, \mu\Omega \, \text{cm}\) (Preising 2023) \\ \hline
\(\boldsymbol{\nu}_e \approx \sqrt{m_i/m_e} \, \nu\) & \(1.5 \times 10^{-6} \sqrt{m_i/m_e} \, \text{m}^2/\text{s}\) (de Wijs 1998) & \(0.4 \times 10^{-6} \sqrt{m_i/m_e} \, \text{m}^2/\text{s}\) (Preising 2023) \\ \hline
\({\nabla} \times \boldsymbol{B}\) & \(10^{-13} \, \text{T/m}\) & \(10^{-14} \, \text{T/m}\) \\ \hline
\rowcolor[gray]{0.7} 
\textcolor{white}{\(\boldsymbol{L}\)} & \textcolor{white}{\(10^{-3} \, \text{m}\)} & \textcolor{white}{\(10 \, \text{m}\)} \\ \hline
\rowcolor[HTML]{1F4E79} 
\textcolor{white}{\(\boldsymbol{r}_i\)} & \textcolor{white}{\(10^{-5} \, \text{m}\)} & \textcolor{white}{\(10^{-4} \, \text{m}\)} \\ \hline
\end{tabular}
\caption{Comparison of parameters for Earth’s liquid iron core and Saturn’s metallic hydrogen. $L$ is the spatial turbulence scale at which the resistive term $\eta\bol{J}$ is comparable to the viscous term $m_e\nu_e\Delta\bol{u}/e$.}
\label{tab:comparison}
\end{table}

    Finally, we remark that similar considerations apply to the ratio of the friction force \(\bol{F}_{\gamma}\) and the resistive term \(\eta \bol{J}\), as $\bol{F}_{\gamma}$ and $\Delta\bol{u}$ model the same dissipative force.

\section{Modified Cowling's Theorem}

{Let us} examine how Cowling's theorem changes under the modified  Ohm's law \eqref{GOhm}. It will be sufficient to consider an axially symmetric poloidal magnetic field 
\eq{
\bol{B}=\nabla\Psi\cp\nabla\varphi,\label{B}
}
where $\Psi\lr{r,z,t}$ denotes the flux function, and an axially symmetric velocity field
\eq{
\bol{u}=\nabla\Theta\cp\nabla \varphi+g\nabla\varphi,
}
with $\Theta\lr{r,z,t}$ and $g\lr{r,z,t}$ single-valued functions. 
For simplicity, {we hereafter} assume $\eta_{\gamma}>0$ to be constant. 
Using \eqref{GOhm}, the induction equation for the magnetic field $\bol{B}$ reads as
\eq{
\frac{\p\bol{B}}{\p t}=\nabla\cp\lr{\bol{u}\cp\bol{B}-\eta_{\gamma}\bol{J}+\gamma\bol{u}}-\gamma\Omega\nabla z. 
}
Substituting \eqref{B}, we have
\eq{
\nabla\cp\lr{\frac{\p\Psi}{\p t}\nabla\varphi}=&\nabla\cp\lrc{-\lr{\nabla\Psi\cdot\bol{u}}\nabla\varphi+\frac{g}{r^2}\nabla\Psi+\frac{\eta_{\gamma}}{\mu_0}\lrs{r\frac{\p}{\p r}\lr{\frac{1}{r}\frac{\p\Psi}{\p r}}+\frac{\p^2\Psi}{\p z^2}}\nabla\varphi}\\
&+\gamma\nabla\cp\lr{\nabla\Theta\cp\nabla\varphi+g\nabla\varphi-\frac{1}{2}\Omega r^2\nabla\varphi}.
}
Considering a toroidal volume $T\subset\mathbb{R}^3$ as domain for the induction equation, we can remove the curl operator by introducing a potential $\Phi\lr{r,z,t}$  such that 
\eq{
\frac{\p\Psi}{\p t}\nabla\varphi=&\nabla\Phi+\lrs{\gamma g-\frac{\gamma}{2}\Omega r^2-\lr{\nabla\Psi\cdot\bol{u}}+\frac{\eta_{\gamma}}{\mu_0}\lr{
\Delta\Psi-2\nabla\Psi\cdot\nabla\log r
%\frac{\p^2\Psi}{\p r^2}-\frac{1}{r}\frac{\p\Psi}{\p r}+\frac{\p^2\Psi}{\p z^2}}
}}\nabla\varphi\\&+\frac{g}{r^2}\nabla\Psi+\gamma\nabla\Theta\cp\nabla\varphi.
}
The toroidal component of this equation reads as
\eq{
\frac{\p\Psi}{\p t}=
\gamma g-\frac{\gamma}{2}\Omega r^2-{\nabla\Psi\cdot\bol{u}}+\frac{\eta_{\gamma}}{\mu_0}
\lr{\Delta\Psi-2\nabla\Psi\cdot\nabla\log r}. \label{eq46}
%\lr{\frac{\p^2\Psi}{\p r^2}-\frac{1}{r}\frac{\p\Psi}{\p r}+\frac{\p^2\Psi}{\p z^2}}.
}
{A key observation is that this governing equation allows for a zero magnetic field solution whenever the toroidal velocity equals the planetary rotational speed, specifically when \( g = {\Omega r^2}/{2} \). Hence, the system is not frictionally driven, allowing for spontaneous dynamo action.}  

Next, we evaluate the integral
\eq{
\frac{1}{2}\frac{d}{dt}\Norm{\Psi}_{T}^2=\frac{1}{2}\frac{d}{dt}
\int_{T}\Psi^2\,dV, 
}
where %we introduced spherical coordinates $\lr{R,\theta,\varphi}$ and 
$\Norm{\cdot}_{T}$ denotes the standard $L^2\lr{T}$ norm. We have
\eq{
\frac{1}{2}&\frac{d}{dt}\Norm{\Psi}_{T}^2=\int_{T}\Psi\lrs{\gamma g-\frac{\gamma}{2}\Omega r^2-\nabla\Psi\cdot\bol{u}+\frac{\eta_{\gamma}}{\mu_0}\lr{\Delta\Psi-2\nabla\Psi\cdot\nabla\log r}}\,dV.
}
Noting that $\Delta\log r=0$, and integrating by parts under the boundary conditions
\sys{
&\Psi=0~~~~{\rm on}~~\p T,\\ 
&\bol{u}\cdot\bol{n}=0~~~~{\rm on}~~\p T,
}{bcs}
where %$\Psi_0\in\mathbb{R}$ and  
$\p T$ denotes the boundary of $T$ with unit outward normal $\bol{n}$, it follows that 
\eq{
\frac{1}{2}\frac{d}{dt}\Norm{\Psi}_{T}^2=%\int_{T}\Psi\lrc{\gamma\lr{g-\frac{1}{2}\Omega r^2}+\eta_{\gamma}\lrs{\nabla\cdot\lr{\Psi\nabla\Psi}-\abs{\nabla\Psi}^2}}\,dV\\%-\eta_{\gamma}\Psi_0^2\int_T\Delta\log r\,dV\\ 
\gamma\int_T\Psi\lr{g-\frac{1}{2}\Omega r^2}\,dV%+\eta_{\gamma}\Psi_0\int_{\p T}\nabla\Psi\cdot\bol{n}\,dS
-\frac{\eta_{\gamma}}{\mu_0}\Norm{\nabla\Psi}^2_T.
}
%Without loss of generality, we may assume $\Psi\geq 0$ in $T$ (if  $\Psi$ is a well-behaved function of space and time, it attains a minimum  $\Psi_m$, and we may define a new  flux function $\Psi'=\Psi+\abs{\Psi_m}\geq 0$). 
%Then, for a sufficiently large toroidal flow $g$, 
For a suitable choice of the toroidal flow $g$, the integrand of the first term on the right-hand side becomes positive, 
%the first-term on the right-hand side becomes positive, 
indicating the possibility of dynamo action. 
If the system achieves an equilibrium, defining $\Psi_{\infty}=\lim_{t\rightarrow+\infty}\Psi$ and $g_{\infty}=\lim_{t\rightarrow+\infty}g$, we find that
\eq{
\Norm{\nabla\Psi_{\infty}}^2_T=\frac{\gamma\mu_0}{\eta_{\gamma}}\int_T\Psi_{\infty}\lr{g_{\infty}-\frac{1}{2}\Omega r^2}\,dV.\label{dPsiL2}
}
Evidently, for a suitable equilibrium toroidal flow $g_{\infty}$ in $T$, the right-hand side becomes positive, and this equality admits non-trivial solutions $\nabla\Psi_{\infty}\neq \bol{0}$ in $T$ {(see below for more details)}. 
Note that, however, the right-hand side of eq. \eqref{dPsiL2} vanishes as soon as the restoring force \eqref{Fgamma} is absent, i.e. $\gamma=0$, recovering Cowling's theorem.  

In $\mathbb{R}^3$, the rate of change in magnetic energy is
\eq{
\frac{1}{2}\frac{d}{dt}\Norm{\bol{B}}_{\mathbb{R}^3}^2=-\mu_0\int_{T}{\bol{E}\cdot\bol{J}}\,dV=\mu_0\int_{T}\lrs{\bol{u}\cp\bol{B}+\gamma\lr{\bol{u}-\bol{v}_R}}\cdot\bol{J}\,dV-\mu_0\eta_{\gamma}\Norm{\bol{J}}^2_{T},
}
where we used Maxwell's equation $\p\bol{B}/\p t=-\nabla\cp\bol{E}$, the fact that $\bol{E}$ and $\bol{B}$ must vanish at infinity, the fact that the current density is confined to $T$,  and equation \eqref{GOhm}. 
Now suppose that the flow $\bol{u}=g\nabla\varphi$ is purely toroidal. Noting that $\mu_0\bol{J}=-\lr{\Delta\Psi-2\nabla\Psi\cdot\nabla\log r}\nabla\varphi$ is purely toroidal, it follows that
\eq{
\frac{1}{2}\frac{d}{dt}\Norm{\bol{B}}_{\mathbb{R}^3}^2=
\mu_0\gamma\int_{T}\lr{{g}-\frac{1}{2}\Omega r^2 }\bol{J}\cdot\nabla\varphi\,dV-\mu_0\eta_{\gamma}\Norm{\bol{J}}^2_T.
}
For example, if the current density has the same orientation as the {planet's} rotation speed $\bol{v}_R$, and the toroidal flow $g$ is large enough, the first term on the right-hand side may exceed the second one, leading to increase in magnetic energy. 
%The equilibrium electric field 
%can be expressed with a single-valued electrostatic potential $\Phi_{\infty}\lr{\bol{x}}$ as 
%$\bol{E}_{\infty}=-\nabla\Phi_{\infty}$. Hence, from \eqref{}
The amplification of the magnetic field is not limited to the toroidal volume $T$. Indeed, in $O$, we have
\eq{
\frac{1}{2}\frac{d}{dt}\Norm{\bol{B}}_O^2=%-\int_O\bol{B}\cdot\nabla\cp\bol{E}\,dV=
-\int_{ O}\nabla\cdot\lr{\bol{E}\cp\bol{B}}\,dV-\mu_0\int_O\bol{E}\cdot\bol{J}\,dV=\int_{\p T}\bol{E}\cp\bol{B}\cdot\bol{n}\,dS,
}
where we used Maxwell's equation $\p\bol{B}/\p t=-\nabla\cp\bol{E}$, the fact that $\bol{E}$ and $\bol{B}$ must vanish at infinity, and the fact that $\bol{J}=\bol{0}$ in $O$ (there is no current outside $T$). The tangential electric field $\bol{E}_t=\bol{n}\cp\lr{\bol{E}\cp\bol{n}}$
 must be continuous across the toroidal boundary $\p T$. 
 Furthermore, since the electric current $\bol{J}$ vanishes in $O$, %and the restoring force $-\gamma\lr{\bol{u}-\bol{v}_R}$ should  vanish at the edge of the liquid shell $\p T$, 
 from \eqref{GOhm} we conclude that
 \eq{
 \bol{E}_t=
 %\bol{B}\cp\bol{u}
 %-\gamma\lr{\bol{u}-\bol{v}_{R}}=
 %-\frac{g}{r^2}\nabla\Psi
 -\gamma\lr{\bol{u}-\bol{v}_{R}}~~~~{\rm on}~~\p T,
 }
 where we used the fact that $\bol{u}\cdot\bol{n}=0$ on $\p T$ (recall \eqref{bcs}). 
 Now suppose that at the time $t=0$ the initial seed electric current $\bol{J}_0=\bol{J}\lr{\bol{x},0}\neq\bol{0}$ is positively oriented along the toroidal direction $\varphi$.   Then, there is some time interval in which $\nabla\Psi$ points toward the center of $T$.  
 %On the other hand, since $\Psi$ is constant on $\p T$, and the magnetic flux is stronger toward the center of $T$, 
More precisely, the unit outward normal on $\p T$ is given by $\bol{n}=-\nabla\Psi/\abs{\nabla\Psi}$, leading to
 \eq{
 \frac{1}{2}\frac{d}{dt}\Norm{\bol{B}}_O^2=\gamma\int_{\p T}\lr{\frac{g}{r^2}-\frac{\Omega}{2}}\abs{\nabla\Psi}\,dS.\label{B2O}
 }
This quantity becomes positive for sufficiently large $g> 0$, and vanishes when $\gamma=0$. 
From this equation, we also see that an  equilibrium solution is given by  $g_{\infty}=r^2\Omega/2$ on $\p T$, i.e. 
\eq{
\lr{\bol{u}_{\infty}-\bol{v}_R}\cdot\nabla\varphi=0~~~~{\rm on}~~\p T,\label{Equ}}
where $\bol{u}_{\infty}=\lim_{t\rightarrow+\infty}\bol{u}$. 

For completeness, it is useful to show  how Cowling's theorem is modified if the modified Ohm's law \eqref{GOhm2} involving viscosity is used in place of \eqref{GOhm}. For simplicity, we consider a purely toroidal flow $\bol{u}=g\nabla\varphi$. We have
\eq{
\frac{1}{2}\frac{d}{dt}\Norm{\bol{B}}^2_{\mathbb{R}^3}=\frac{\mu_0 m_e\nu_e}{e}\int_{T}\lr{2\nabla g\cdot\nabla\log r-\Delta g}\bol{J}\cdot\nabla\varphi\,dV-\mu_0\eta\Norm{\bol{J}}^2_T.
}
We thus see that the first term on the right-hand side can be positive for a suitable toroidal flow $g$, indicating the possibility of dynamo action. 
A similar calculation to eq. \eqref{B2O} shows that magnetic field growth is not limited to $T$, 
\eq{
\frac{1}{2}\frac{d}{dt}\Norm{\bol{B}}^2_{O}=\frac{m_e\nu_e}{e}\int_{\p T}r^{-2}\lr{2\nabla g\cdot\nabla\log r-\Delta g}{\abs{\nabla\Psi}}\,dS.
}

\section{Modified Neutral Point Argument}

Cowling's theorem is related to the so-called neutral point argument. 
Let's see how the neutral point argument changes in the present setting. 
Suppose that the equilibrium flux function $\Psi_{\infty}$ is a well-behaved  non-constant function in $\mathbb{R}^3$. 
%, and that a constant is added so that $\Psi_{\infty}$ vanishes on the symmetry axis $r=0$. Since $\abs{\bol{B}=\abs{\nabla\Psi}r^{-1}}$, $\p\Psi/\p r\sim r^{1+\epsilon}$ for some $\epsilon>0$, implying $\Psi_{\infty}\sim r^{2+\epsilon}$v.  
%Hence, if it 
%is a well-behaved and non-constant function of space, 
Then, by symmetry, it will attain local maximima or minimima somewhere in $\mathbb{R}^3$. %(because the value on the symmetry axis must match the value at infinity). 
At these neutral points, $\nabla\Psi_{\infty}$ and  $\bol{B}_{\infty}=\nabla\Psi_{\infty}\cp\nabla\varphi$ vanish. Let $N$ denote a neutral point, and consider a circle $C$ of radius $\epsilon\sim
%\abs{\bol{B}_{\infty}\lr{N+\epsilon\bol{\delta}}-\bol{B}_{\infty}\lr{N}}=
\abs{\bol{B}_{\infty}\lr{N+\epsilon\bol{\delta}}}$, where $\bol{\delta}$ is some unit vector. Let $\Sigma_C$ denote a surface in the $\lr{r,z}$-plane, with normal $\bol{n}_C=r\nabla\varphi$,   enclosed by $C$ and such that $N\in \Sigma_C$. 
Using Ohm's law \eqref{Ohm}, Stokes' theorem, and noting that, at equilibrium, the electric field $\bol{E}_{\infty}=-\nabla\Phi_{\infty}$ is given as the gradient of a single-valued potential $\Phi_{\infty}\lr{{r,z}}$, 
we have
\eq{
0=\int_{\Sigma_C}\bol{E}_{\infty}\cdot\bol{n}_C\,dS=%\int_{\Sigma_C}\lr{\bol{B}_{\infty}\cp\bol{u}_{\infty}+\eta\bol{J}_{\infty}}\cdot\bol{n}_C\, dS=
\int_{\Sigma_C}\bol{B}_{\infty}\cp\bol{u}_{\infty}\cdot\bol{n}_C \,dS+\frac{\eta}{\mu_0}\int_C\bol{B}_{\infty}\cdot d\bol{l}. 
}
However, by construction $\Sigma_C=\pi\epsilon^2$ while $C=2\pi\epsilon$. 
In proximity of $N$, the equation above can thus be satisfied only if 
$\frac{\mu_0}{\eta}\bol{u}_{\infty}\sim {\epsilon^{-1}}$, 
implying that $\bol{u}_{\infty}$ is singular at $N$ for non-vanishing and bounded coefficients $\eta$ and $\mu_0$. 
If the same argument is applied to the modified Ohm's law \eqref{GOhm}, one obtains
\eq{
0=\int_{\Sigma_C}\lrs{\bol{B}_{\infty}\cp\bol{u}_{\infty}-\gamma\lr{\bol{u}-\bol{v}_R}}\cdot\bol{n}_C \,dS+\frac{\eta_{\gamma}}{\mu_0}\int_C\bol{B}_{\infty}\cdot d\bol{l}. 
}
Now the second and third terms on the right-hand side are both second order terms in $\epsilon$, and the neutral point argument does not apply. 

We conclude this section by noting that modifying Ohm’s law only by retaining the Hall effect does not suffice to invalidate the neutral point argument. This is because \( \boldsymbol{J}_{\infty} \times \boldsymbol{B}_{\infty} \cdot \boldsymbol{n}_C = 0 \), given that \( \boldsymbol{J}_{\infty} \times \boldsymbol{n}_C = \boldsymbol{0} \).

\section{Axially Symmetric Poloidal Equilibrium Configurations}
Let's construct axially symmetric poloidal equilibrium configurations, i.e. steady solutions of the induction equation such that 
\sys{
&\bol{B}_{\infty}=\nabla\Psi_{\infty}\cp\nabla\varphi,\\
&\bol{J}_{\infty}=-\frac{1}{\mu_0}\lr{\Delta\Psi_{\infty}-2\nabla\Psi_{\infty}\cdot\nabla\log r}\nabla\varphi,\\
&\bol{E}_{\infty}=-\nabla\Phi_{\infty},\\
&\bol{u}_{\infty}=\nabla\Theta_{\infty}\cp\nabla\varphi+g_{\infty}\nabla\varphi.
}{PolEq}  
Recalling the modified Ohm's law \eqref{GOhm}, the equilibrium equation is
\eq{
-\nabla\Phi_{\infty}=&\lr{\nabla\Theta_{\infty}\cp\nabla\varphi\cdot\nabla\Psi_{\infty}-\frac{\eta_{\gamma}}{\mu_0}\Delta\Psi_{\infty}+2\frac{\eta_\gamma}{\mu_0}\nabla\Psi_{\infty}\cdot\nabla\log r-\gamma g_{\infty}+\frac{1}{2}\gamma\Omega r^2}\nabla\varphi\\&-\frac{g_{\infty}}{r^2}\nabla\Psi_{\infty}-\gamma\nabla\Theta_{\infty}\cp\nabla\varphi.
}
The poloidal component of $\bol{u}_{\infty}$ can be eliminated by setting $\Theta_{\infty}={\rm constant}$. 
On the other hand, the electrostatic potential must be single-valued, leading to $g_{\infty}=r^2 df\lr{\Psi_{\infty}}/d\Psi_{\infty}$ and  $\Phi_{\infty}=f\lr{\Psi_{\infty}}$. 
The equilibium flux function $\Psi_{\infty}$ is then determined by a Poisson equation. 
The full set of equilibrium equations is
\sys{
&\Delta\Psi_{\infty}=2\nabla\Psi_{\infty}\cdot\nabla\log r-\frac{\gamma\mu_0}{\eta_{\gamma}} r^2\lr{\frac{df\lr{\Psi_{\infty}}}{d\Psi_{\infty}}-\frac{\Omega}{2}}~~~~{\rm in}~~T,\label{etaJnu}\\
&\Phi_{\infty}=f\lr{\Psi_{\infty}}~~~~{\rm in}~~T,\label{EBxu}\\
&g_{\infty}=r^2\frac{df\lr{\Psi_{\infty}}}{d\Psi_{\infty}}~~~~{\rm in}~~T,\\
&\Theta_{\infty}={\rm constant}~~~~{\rm in}~~T,\\
&\Psi_{\infty}=0~~~~{\rm on}~~\p T,\\
&\frac{df\lr{\Psi_{\infty}}}{d\Psi_{\infty}}=\frac{\Omega}{2}~~~~{\rm on}~~\p T\label{bcg}.
}{PolEq2}
%Note that, 
{For} given $f\lr{\Psi_{\infty}}$, this system admits nontrivial regular solutions in $T$ thanks to the source term proportional to $\gamma$ in the first equation. 
{The result reveals that the} electric field $\bol{E}_{\infty}$ is balanced by the ideal convective term $\bol{B}_{\infty}\cp\bol{u}_{\infty}$ (see eq. \eqref{EBxu}), while the resistive term $\eta_{\gamma}\bol{J}_{\infty}$ is balanced by the friction term $-\gamma\lr{\bol{u}_{\infty}-\bol{v}_{R}}$ as shown in eq. {\eqref{etaJnu}}. 
This type of force balance is consistent with kinetic theory of plasmas, which informs us that equilibria of the Boltzmann equation usually  
belong to the kernel of the collision operator, i.e. in a steady state  ideal terms and dissipative terms vanish independently (see e.g. \cite{Sato24} and references therein). Indeed, MHD and related 2-fluid theories are reduced models originating from kinetic theory.  
{As previously noted, zero magnetic field solutions are permitted when $g_{\infty} = \frac{r^2 \Omega}{2}$ in $T$. Additionally, it is important to observe that the toroidal component of the fluid velocity, $g_{\infty} \nabla \varphi$, can deviate significantly from the planetary rotation speed, $\bol{v}_R = \frac{1}{2} \Omega r^2 \nabla \varphi$, within $T$ when a non-zero magnetic field is present (see eq. \eqref{etaJnu}).}

We also remark that $\bol{J}_{\infty}$ vanishes on $\p T$ due to the boundary condition \eqref{bcg} coming from equation \eqref{Equ}. 
Hence, for any $\bol{x}\in\mathbb{R}^3$ the (continuous)  magnetic field vanishing at infinity and corresponding to the current distribution resulting from system \eqref{PolEq2} can be obtained through the Biot-Savart integral,  
\eq{
\bol{B}_{\infty}\lr{\bol{x}}=\frac{\mu_0}{4\pi}\int_{T}\frac{\bol{J}_{\infty}\lr{\bol{x}'}\cp\lr{\bol{x}-\bol{x}'}}{\abs{\bol{x}-\bol{x}'}^3}\,dV'.\label{Binfty}
}
%where $\mu_0$ is the vacuum permeability (we neglect effects associated with magnetization). 
The properties of the Biot-Savart integral \eqref{Binfty}, as well as a generalization to bounded domains, can be found in \cite{Enciso18}.  
We remark that since the current density $\bol{J}_{\infty}$ is an axially symmetric toroidal flow, it follows that in $O$ the equilibrium magnetic field $\bol{B}_{\infty}$ is an axially symmetric dipole-type magnetic field.

\section{Concluding Remarks}

%In conclusion, this work significantly enhances our understanding of the long-term %behavior of geophysical and astrophysical dynamos. 
% **THIS WOULD BE NICER FOR THE COVER LETTER, BUT MAYBE NOT HERE IN THE MAIN TEXT?**

We have proposed a novel dynamo model that incorporates a modified Ohm's law within a toroidal volume. The geometry of this toroidal volume, inspired by analogies with fusion plasma physics, can be viewed as a nearly spherical dynamo-active region enclosed by the liquid shell surrounding the planetary core.
It is important to note that, whether the system involves liquid iron in rocky planets or ionized hydrogen in gas giants, its electromagnetic behavior is fundamentally governed by microscopic interactions between ions and electrons, thereby justifying the use of a two-fluid model.
In this framework, Ohm's law is modified to account for the dissipative force associated with electron collisions that tends to align macroscopic flows with {the planet's} rotation speed. 
This force is modeled either as a frictional damping term, or as a viscous force.   
Specifically, the two models examined are as follows. The first model introduces a frictional (damping) force that tends to align  the fluid velocity with the planetary rotation speed. The second model treats the dissipative force as viscous dissipation, under the assumption that the rotating fluid shell adheres to a Navier-Stokes reduction of the underlying kinetic equations.
Our analysis shows that this dissipative force can be comparable to the conventional resistive force associated with electron-ion collisions in the standard formulation of Ohm's law. Therefore, it should not be neglected.

%In both cases, the resulting dynamo model successfully generates an axially symmetric dipolar magnetic field driven by an ion flow aligned with the steady electric current density. 

The presented approach {indeed} enables the formation of axially-symmetric steady-state magnetic fields, circumventing traditional spherical geometries and anti-dynamo theorems{.
The} key components of our model include: (i) identifying the dissipative force acting on the {fluid} shell, in addition to collisional resistive forces, as a critical factor in the electron momentum equation.
(ii) Modeling the convecting region as a toroidal volume, which allows for non-trivial steady states of the induction equation.
(iii) Focusing on steady states where ideal and dissipative terms balance independently.

Note that the governing equation, e.g. \eqref{eq46}, allows for a zero magnetic field solution when the toroidal velocity matches the planetary rotational speed, indicating that the system is not frictionally driven. This implies that the current model can spontaneously generate a magnetic field purely through fluid motion.
These findings adhere to anti-dynamo theorems: we demonstrate how Cowling's theorem and the neutral point argument are modified by the dissipative force, enabling the existence of non-trivial, axially symmetric equilibrium magnetic fields. 

Our model successfully captures the essence of a dynamo, producing a stable dipolar magnetic field and providing valuable insights into the long-term operation and stability of geophysical and astrophysical dynamos. The model's foundation in partial differential equations ensures broad applicability, including potential generalizations to oscillating magnetic field amplification. 
To determine the precise range of parameters for which the theory is applicable, experimental and numerical validation are necessary.

\section*{Statements and declarations}

\subsection*{Data availability}
Data sharing not applicable to this article as no datasets were generated or analysed during the current study.

\subsection*{Funding}
The research of NS was partially supported by JSPS KAKENHI Grant No.  22H04936 and No. 24K00615. 
KH also acknowledges support from JSPS KAKENHI Grant-in-Aid (B) No.~24K00694. 
%This work was partly supported by MEXT Promotion of Distinctive Joint Research Center Program JPMXP0723833165. 

\subsection*{Competing interests} 
The authors have no competing interests to declare that are relevant to the content of this article.


\begin{thebibliography}{}

\bibitem{Moffatt78}
H. K. Moffatt, \textit{Laminar dynamo theory}, In \textit{Magnetic field generation in electrically conducting fluids}, ed. G. K. Batchelor and J. W. Miles, pp. 108-144, Cambridge University Press (1978).

\bibitem{Desjardins07}
B. Desjardins, E. Dormy, A. Gilbert, and M. Proctor, \textit{Introduction to self-excited dynamo action}, In \textit{Mathematical aspects of natural dynamos}, ed. E. Dormy and A. M. Soward, pp. 3-57, CRC Press (2007).

\bibitem{Tobias21}
S. M. Tobias, \textit{The turbulent dynamo}, J. Fluid Mech., \textbf{912}, PI, 1-76 (2021).

\bibitem{Larmor19}
J. Larmor, \textit{How could a rotating body such as the sun become a magnet?}, Brit. Assoc. Adv. Sci. Rep., 159-160 (1919).

\bibitem{Cowling33}
T. G. Cowling, \textit{The magnetic field of sunspots}, Mon. Not. R. Astr. Soc., \textbf{94}, 1, 39-48 (1933).

\bibitem{Krause80}
F. Krause, and K. H. Raedler, In \textit{Mean-field magnetohydrodynamics and dynamo theory}, Pergamon Press (1980).

\bibitem{Brag76}
S. I. Braginsky, \textit{On the nearly axially-symmetrical model of the hydromagnetic dynamo of the earth}, Phys. Earth Planet. Int., \textbf{11}, 3, 191-199 (1976).

\bibitem{Backus58}
G. Backus, \textit{A class of self-sustaining dissipative spherical dynamos}, Ann. Phys., \textbf{4}, 372-447 (1958).

\bibitem{Ponomarenko73}
Y. B. Ponomarenko, \textit{Theory of the hydromagnetic generator}, J. Appl. Mech. Tech. Phys., \textbf{14}, 775-778 (1973).

\bibitem{Ivers84}
D. J. Ivers, and R. W. James, \textit{Axisymmetric antidynamo theorems in compressible nonuniform conducting fluids}, Phil. Trans. R. Soc. Lond. A, \textbf{312}, 179-218 (1984).

\bibitem{Nunez96}
M. Núñez, \textit{Axisymmetric dynamo solutions}, SIAM Rev., \textbf{38}, 553-564 (1996).

\bibitem{Kaiser14}
R. Kaiser, and A. Tilgner, \textit{The axisymmetric antidynamo theorem revisited}, SIAM J. Appl. Math., \textbf{74}, 571-597 (2014).

\bibitem{Lortz89}
D. Lortz, \textit{Axisymmetric dynamo solutions}, Z. Naturforsch., \textbf{44a}, 1041-1045 (1989).

\bibitem{Plunian20}
F. Plunian, and T. Alboussière, \textit{Axisymmetric dynamo action is possible with anisotropic conductivity}, Phys. Rev. Res., \textbf{2}, 013321 (2020).

\bibitem{Yadav22}
R. K. Yadav, H. Cao, and J. Bloxham, \textit{A dynamo simulation generating Saturn-like small magnetic dipole tilt}, Geophys. Res. Lett., \textbf{49}, e2021GL097280 (2022).

\bibitem{Freid}
J. P. Freidberg, \textit{The ideal MHD model}, In \textit{Ideal MHD}, p. 25, Cambridge University Press (2014).

\bibitem{Wesson04}
J. Wesson, \textit{Equilibrium}, In \textit{Tokamaks}, 3rd ed., pp. 104-110, Oxford University Press (2004).

\bibitem{Helander14}
P. Helander, \textit{Theory of plasma confinement in non-axisymmetric magnetic fields}, Rep. Prog. Phys., \textbf{77}, 087001, 1-35 (2014).

\bibitem{Eisenberg79}
M. Eisenberg, and R. Guy, \textit{A proof of the hairy ball theorem}, Am. Math. Mon., \textbf{86}, 571-574 (1979).



\bibitem{Sato24}
N. Sato, and P. J. Morrison, \textit{A collision operator for describing dissipation in noncanonical phase space}, Fundamental Plasma Physics, \textbf{10}, 100054 (2024).

\bibitem{Cao23}
{ H. Cao, M. K. Dougherty, G. J. Hunt, 
E. J. Bunce, U. R. Christensen, K. K. Khurana, and M. G. Kivelson}, \ti{Saturn's magnetic field at unprecedented detail achieved by Cassini's close encounters} in 
Cassini at Saturn: The Grand Finale, 15pp. 
To be published by Cambridge University Press (2023).

\bibitem{Anderson12}
%\kumiko{
B. J. Anderson, C. L. Johnson, H. Korth, R. M. Winslow, J. E. Borovsky, M. E. Purucker, J. A. Slavin, S. C. Solomon, M. T. Zuber, and R. L. McNutt Jr., \ti{Low-degree structure in Mercury's planetary magnetic field},  
{ J. Geophys. Res. } {\bf 117}, E00L12 (17pp) (2012).
%}

% {\sc Cao, H., Dougherty, M.K., Hunt, G.J., Bunce, E.J., Christensen, U.R., Khurana, K.K. and Kivelson, M.G.} 2023 {Saturn's magnetic field at unprecedented detail achieved by Cassini's close encounters},  {In \it Cassini at Saturn: The Grand Finale}, %{ed K. Baines, M. Flasar, N. Krupp, and T. Stallard},
% 15pp. To be published by Cambridge University Press. 
 %Also available as arXiv:2301.02756 [astro-ph.EP].
 %%at \url{https://doi.org/10.48550/arXiv.2301.02756}.

\bibitem{Stevenson80}
D. J. Stevenson, \textit{Saturn's Luminosity and Magnetism}, Science, \textbf{208}, 4445, 746-748 (1980).

\bibitem{Christensen06}
%\kumiko{
U. R. Christensen,  
\ti{A deep dynamo generating Mercury's magnetic field}, 
{ Nature} {\bf 444}, pp. 1056-1058 (2006).
%}

\bibitem{Ash}
 {N. W. Ashcroft and N. D. Mermin,} \ti{The Drude theory of metals},  {In Solid State Physics},   Saunders College Publishing, pp. 1-27 (1976). 

\bibitem{Fitz} 
{R. Fitzpatrick}, \ti{MHD equations}, 
{In Plasma Physics: An Introduction}, CRC Press,  
pp. 107-108. (2015). 

\bibitem{Ach}
{M. Acheritogaray, P. Degond, A. Frouvelle, and J. G. Liu,} \ti{Kinetic formulation and global existence for the Hall-magneto-hydrodynamics system}, { Kinet. Relat. Models} {\bf 4}, 4, pp. 901-918 (2011).  
%M. Acheritogaray, P. Degond, A. Frouvelle, and J. G. Liu, “Kinetic formulation and global existence for the Hall-magneto-hydrodynamics system,” Kinet. Relat. Models 4(4), 901–918 (2011).

\bibitem{Abd}
{H. M. Abdelhamid, Y. Kawazura, and Z. Yoshida,} 
 \ti{Hamiltonian formalism of extended magnetohydrodynamics}, { J. Phys. A: Math. Theor.} {\bf 48}, 235502 (2015). 
%H. M. Abdelhamid, Y. Kawazura, and Z. Yoshida, “Hamiltonian formalism of extended magnetohydrodynamics,” J. Phys. A: Math. Theor. 48(23), 235502 (2015).



\bibitem{Minnini} 
{P. D. Minnini, D. O. G\'omez, and S. M. Mahajan,}  \ti{Dynamo action in magnetohydrodynamics and Hall-magnetohydrodynamics}, 
{ The Astrophysical Journal} {\bf 587}, pp. 472-481 (2003).

\bibitem{Lingam}
M. Lingam and S. M. and Mahajan,  \ti{Modelling astrophysical outflows via the unified dynamo-reverse dynamo mechanism}, 
{ Monthly Notices of the Royal Astronomical Society} {\bf 449}, pp. L36-L40 (2015). 




 

\bibitem{Ohta16}
K. Ohta, Y. Kuwayama, K. Hirose, K. Shimizu, and Y. Ohishi, \textit{Experimental determination of the electrical resistivity of iron at Earth's core conditions}, Nature, \textbf{534}, 95-98 (2016).

\bibitem{Wijs98}
G. A. de Wijs, G. Kresse, L. Vocadlo, D. Dobson, D. Alfè, M. J. Gillan, and G. D. Price, \textit{The viscosity of liquid iron at the physical conditions of the Earth's core}, Nature, \textbf{392}, 805-807 (1998).

\bibitem{Preising23}
M. Preising, M. French, C. Mankovich, F. Soubiran, and R. Redmer, \textit{Material properties of Saturn's interior from ab initio simulations}, The Astrophysical Journal Supplement Series, \textbf{269}, 47 (2023).










\bibitem{Enciso18}
A. Enciso, M. A. Garcia-Ferrero, and D. Peralta-Salas, \textit{The Biot-Savart operator of a bounded domain}, Journal de Mathématiques Pures et Appliquées, \textbf{119}, 85-113 (2018).










%\bibitem{Jones10}
%C. A. Jones, M. J. Thmpson, and S. M.  Tobias, \ti{The Solar Dynamo}, Space Sci. Rev., {\bf 152}, 591-616 (2010).





















%\bibitem{Roberts71}
%P. H. Roberts, and M. Stix, \textit{The Turbulent Dynamo: A Translation of a Series of Papers by F. Krause, K.-H Radler, and M. Steenbeck}, NCAR Technical Notes, IA-60 (1971).


\end{thebibliography}
\end{document}